%% file: main.tex
\pdfoutput=1
\documentclass[sigconf,nonacm]{acmart}
\def\BibTeX{{\rm B\kern-.05em{\sc i\kern-.025em b}\kern-.08emT\kern-.1667em\lower.7ex\hbox{E}\kern-.125emX}}

\usepackage{booktabs} % For formal tables
\usepackage{graphicx}
\usepackage{xcolor}
\usepackage{float}
\usepackage{caption}
\usepackage{xspace}
\newcommand{\Sitara}{Sitara}
\begin{document}
\title{\Sitara: Spectrum Measurement Goes Mobile Through Crowd-sourcing}
%\title{pocketSDR: Spectrum Monitoring System Goes Mobile Through Crowd-sourcing}
%\title{Crowd-sourced Spectrum Monitoring Goes Mobile with pocketSDR}%
%\title{Put on Your Shoes! Crowd-sourced Spectrum Monitoring Goes Mobile with pocketSDR}%

\author{Phillip Smith}
\affiliation{%
\institution{University of Utah}}
\email{phillip.smith@utah.edu}

\author{Anh Luong}
\affiliation{%
\institution{Carnegie Mellon University}}
\email{anhluong@cmu.edu}

\author{Shamik Sarkar}
\affiliation{%
\institution{University of Utah}}
\email{shamik.sarkar@utah.edu}

\author{Harsimran Singh}
\affiliation{%
\institution{University of Utah}}
\email{hsing4@cs.utah.edu}

\author{Neal Patwari}
\affiliation{%
\institution{University of Utah}}
\email{npatwari@ece.utah.edu}

\author{Sneha Kasera}
\affiliation{%
\institution{University of Utah}}
\email{kasera@cs.utah.edu}

\author{Kurt Derr}
\affiliation{%
\institution{Idaho National Labs}}
\email{kurt.derr@inl.gov}

\author{Samuel Ramirez}
\affiliation{%
\institution{Idaho National Labs}}
\email{samuel.ramirez@inl.gov}




%\thispagestyle{empty}
% Software-defined radios (SDRs) are often used in the experimental evaluation of next-generation wireless technologies.  While crowd-sourced spectrum monitoring is an important component of future spectrum-agile technologies, there is no clear way to test it in the real world, i.e., with hundreds of users each with an SDR in their pocket participating in RF experiments controlled by, and data uploaded to, the cloud.  Current fully functional SDRs are bulky, with components connected via wires, and last at most hours on a single battery charge.  To address the needs of such experiments, we design and develop a compact, portable, untethered, and inexpensive SDR we call \emph{\Sitara}.

\begin{abstract}
Software-defined radios (SDRs) are often used in the experimental evaluation of next-generation wireless technologies. While crowd-sourced spectrum monitoring is an important component of future spectrum-agile technologies, there is no clear way to test it in the real world, i.e., with hundreds of users each with an SDR in their pocket participating in RF experiments controlled by, and data uploaded to, the cloud. Current fully functional SDRs are bulky, with components connected via wires, and last at most hours on a single battery charge. To address the needs of such experiments, we design and develop a compact, portable, untethered, and inexpensive SDR we call \emph{\Sitara}. Our SDR interfaces with a mobile device over Bluetooth 5 and can function standalone or as a client to a central command and control server. The {\Sitara} offers true portability: it operates up to one week on battery power, requires no external wired connections and occupies a footprint smaller than a credit card. It transmits and receives common waveforms, uploads IQ samples or processed receiver data through a mobile device to a server for remote processing and performs spectrum sensing functions. Multiple {\Sitara}s form a distributed system capable of conducting experiments in wireless networking and communication in addition to RF monitoring and sensing activities. In this paper, we describe our design, evaluate our solution, present experimental results from multi-sensor deployments and discuss the value of this system in future experimentation.
\end{abstract}

\maketitle

\input{parts/introduction}
\input{parts/overview}
\input{parts/implementation}

\input{parts/evaluation}
\input{parts/results}

\input{parts/related_work}

\input{parts/conclusion}

{\footnotesize \bibliographystyle{acm}
\bibliography{references}

\end{document}

%% file: parts/introduction.tex
\section{Introduction}
\label{intro}
Future mobile wireless advancements will continue a trend of increasing densification, distribution and coordination, and spectrum-agile operation~\cite{shafi2017,wang2014cellular,sharma2017dynamic}. The performance of these new technologies depends not only on the mobility of individual users with respect to base stations, but users' mobility with respect to each other. Ideally, to quantify performance experimentally, one would run a large-scale distributed wireless experiment with tens or hundreds of software-defined radios (SDR) programmed to deploy/test a new technology, while individual volunteers each carry these SDRs with them during their normal daily activities.  Such an experiment would allow technologies to be tested with users' real-world mobilities, including temporal, spatial, and person-to-person correlations, rather than in artificial testbed or simulation environments that implicitly or explicitly assume independence and stationarity.

For researchers to be able to run such experiments, the SDR must be \emph{truly portable}, so that a volunteer participant is not burdened by the carrying of the device, and in fact does not ordinarily notice it.  Furthermore, the hardware must be \emph{low cost} to enable experiments with hundreds of volunteers.  We define portable to mean capable of operating for extended periods of time without an external power supply, small enough to carry without encumbering the user --- small enough to easily fit into a pocket --- and {\em not tethered} to wires, cables, %antennas, 
or external connectors. Its energy consumption must allow it to last as long as most smartphones so that volunteers can charge it on the same schedule as their mobile device.
In terms of cost, a researcher should be able to purchase a set of 100 on a standard grant, which would translate to around US \$50 or less per device.

Unfortunately, no existing SDR meets the requirements of performing such experiments. Recent products include devices such as the Kickstarter-funded ``portable SDR'' (PSDR)~\cite{ColtonSDR}, RTL-SDR~\cite{rtlSDR}, LimeSDR~\cite{limeSDR} and Ettus USRP radios~\cite{ettus}. While each of these possess useful capabilities, all fall short in one area or another for large scale mobile experiments. The most suitable among these would be the Ettus USRP E312, a battery-powered portable SDR~\cite{e312}. The E312, however, is far too large to fit into a pocket and it must be tethered, i.e., it would still require a cable connection to a mobile device or laptop to provide a control channel. Additionally, the steep price of the E312 would present a practical limit to large scale distributed experiments.  The RTL-SDR is low cost, but requires an external processor, to which it must be connected to by USB cable, and most importantly, has no transmit capability. Most true SDRs, capable of digital processing of RF samples and RX/TX, cost at least US \$100 in the most basic form and quickly reach 10 times that cost for more sophisticated offerings. Such SDRs quickly become cost-prohibitive for large-scale experiments, and are not appreciated by volunteers for their size or tethering requirements.

We also point out that current mobile phones, while packed with cellular, Wi-fi and Bluetooth radios, do not provide the flexibility to make arbitrary changes across layers which researchers want to explore, despite their suitability for some particular applications in localization and sensing \cite{improved_wifi, wireless_localization, fingerprint}. Differences in chipsets, firmware implementation, protocols and carrier-imposed restrictions preclude uniform or arbitrary access and control of the underlying hardware. Even with unrestricted access to the underlying hardware, such operation would disrupt data services and inconvenience the volunteer --- something we wish to avoid.

\subsection{A Novel Software-Defined Radio}
Our contribution is a novel open-source device and cloud framework aimed at enabling large-scale experimental research in mobile dynamic spectrum access, propagation modeling, distributed and coordinated reception, and localization.  Our device, that we call \emph{\Sitara}, is a truly portable software-defined radio. It is especially suited for distributed, crowd-sourced experiments. It is designed to have a battery life of up to a week on a single charge, to be smaller than a credit card, and to cost less than existing fully-featured SDRs.  Our {\Sitara} is convenient for volunteers to carry and is accessible to a broad set of researchers. We anticipate this to be particularly useful for scenarios in which simultaneous, near real-time, geographically distributed narrow-band RF measurements are desired. We will later demonstrate how the {\Sitara} can become a valuable tool, quickly amassing measurement inputs for models and providing insights to inform decisions for wireless research.
%network architectures.

\subsection{Achieving True Portability}
The aim of achieving a compact, cordless, energy efficient device constrains key design decisions. The inconvenience of frequent charging and limited space for batteries make the power requirements of field-programmable gate arrays (FPGAs) commonly used in other SDR solutions unfeasible~\cite{fpga}. For a device to be practical for crowd-sourcing it must also be convenient for volunteers to carry, which means we cannot connect it via cable to their smartphone, and little to no interaction should be required from the volunteer. With the recent availability of Bluetooth 5 devices and the ubiquity of smartphones, we arrive at the solution presented here: a low power transceiver paired with a Bluetooth interface. By pairing with the volunteer's phone, we can piggyback on the phone's WiFi or cellular connection to communicate with a remote server, as well as its location service.

But how can we avoid the large cost and size of most fully-functional SDRs?  We apply a lesson from the RTL-SDR, which re-purposed a mass-produced digital video receiver (the RTL2832U) for its ability to output complex-baseband (IQ) samples.  
We use the Texas Instruments (TI) CC1200 transceiver \cite{cc1200} which, although not designed as an SDR transceiver, has an IQ sample feature as well as transmit capability.
This transceiver supports operation below 1 GHz. The experiments we perform in this paper are in the 902-928 MHz ISM band. Our firmware limits transmission to one of several ISM bands below 1 GHz. 
The {\Sitara} complements the CC1200 with a Nordic Semiconductor nRF52840 system on chip (SoC) and supporting circuitry \cite{nrf}. This SoC contains a single-core ARM microcontroller ($\mu$C) as well as a Bluetooth 5 stack for communication with a mobile device/gateway and a serial peripheral interface (SPI) for communication with the CC1200 transceiver. Any mobile app compatible with the Nordic UART Bluetooth service can interact with the device.  The {\Sitara} supports reading from and writing to arbitrary registers on the CC1200 radio, tuning radio frequency, measuring RSSI, continuously capturing IQ samples, sample capture on carrier-sense, frequency phase lock, and transmission and reception of messages using various modulations.

\begin{figure}[ht]
%\raggedright
\includegraphics[width=8.4cm]{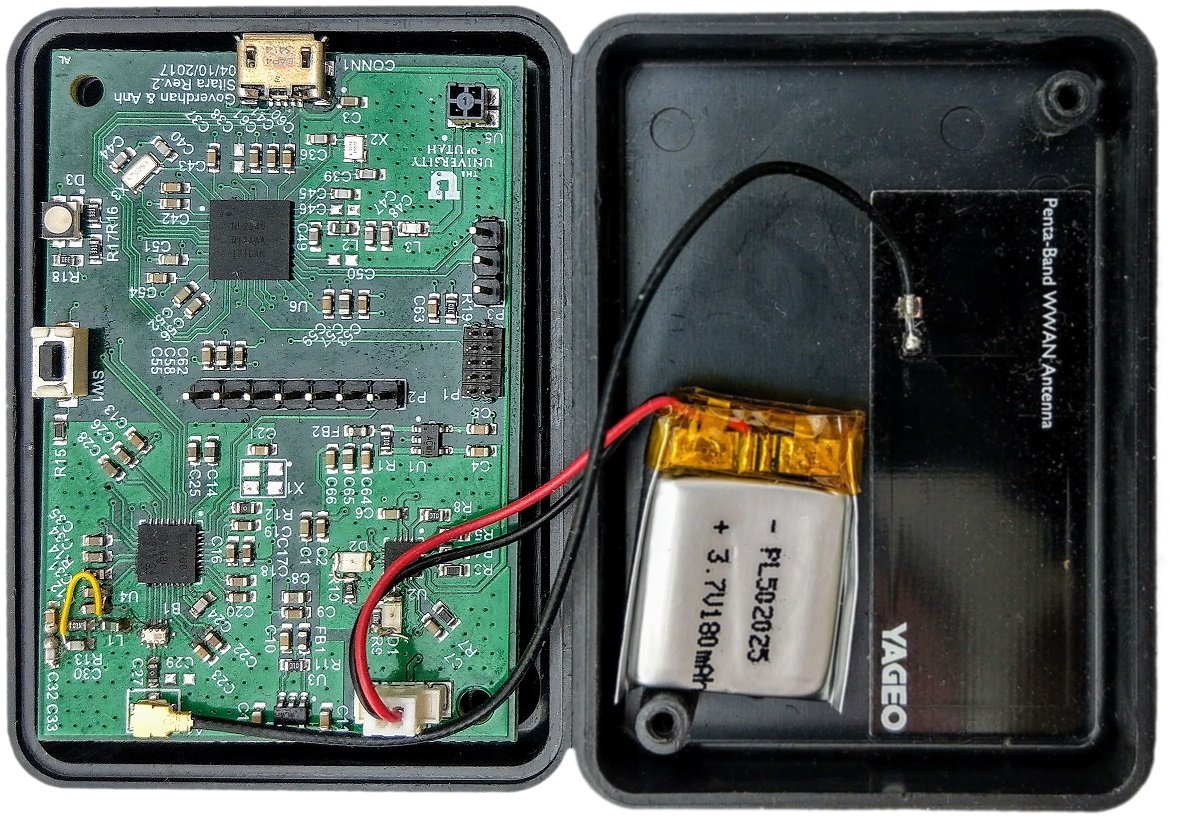}
\caption{{\Sitara} PCB with antenna and rechargable battery housed in an ABS plastic enclosure}
\label{image:sitara}
\end{figure}

\subsection{Balancing Power and Throughput}
For any mobile device, the power budget is always an area of concern. For our application, while targeting low-cost, lower power components, we introduce the challenge of maintaining high sample throughput on hardware that was originally designed for intermittent, bursty operation; continuously operating the $\mu$C alone would deplete our initially specified battery in a matter of hours. We address this problem by designing an architecture that maximizes efficiency by exploiting hardware peripherals to maintain a high data rate while minimizing $\mu$C activity. To prove useful as an SDR, we must maintain a uniform sampling rate for IQ data. This requires solving a difficult problem to achieve a careful coordination between transceiver data acquisition, sample processing and the Bluetooth radio. 

In general terms, we achieve this by minimizing processing overhead in software and optimizing parameters for Bluetooth transmission. Among the Bluetooth features that make this possible are Low-Energy Data Packet Length Extension introduced in Bluetooth 4.2 and the optional 2~Mb/s bit rate, LE 2M PHY, introduced in Bluetooth 5~\cite{blue}. Because the Bluetooth stack is implemented as a ``SoftDevice'', a precompiled binary image, which runs on the single ARM core, there is inherent contention for the $\mu$C's resources. Any timing anomalies occurring while servicing interrupts by the SoftDevice result in a critical fault. Consequently, the SoftDevice must be given interrupt priority, resulting in non-deterministic timing for servicing other interrupts such as sample capture. To overcome this challenge and guarantee uniform sampling, we use the on-chip programmable peripheral interconnect (PPI) together with receive buffers accessed through a direct memory access (DMA) controller to automate necessary tasks, independent of the $\mu$C. To maintain timing uniformity during sampling, we synchronize transfers with a hardware-enabled timer to record the delay caused by any buffer overflows. This allows us to preserve relative symbol timing even if some samples are lost due to an overflow. We further explain this approach and provide more details in Section \ref{section:design}. Our solution allows continuous sample capture over SPI and only requires $\mu$C intervention to rotate between receive buffers. The result of our efforts is a maximum, hardware-limited, continuous sampling rate up to 104 kS/s across the SPI interface and Bluetooth data rates exceeding 1Mb/s. The maximum sampling rate across the SPI interface effectively limits our receiver bandwidth to 52 kHz for IQ sampling.

\subsection{Cloud-based Command and Control Server}
\begin{figure}[ht]
\raggedright
\includegraphics[width=8.4cm]{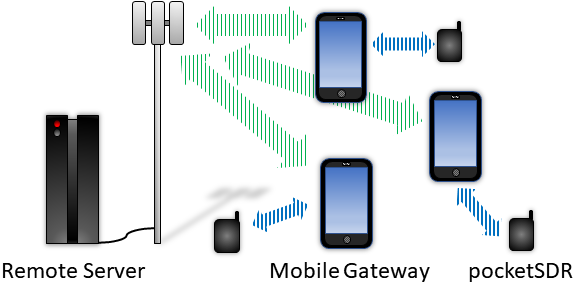}
\caption{{\Sitara} backhaul system includes the Bluetooth connection (blue) between the {\Sitara} and mobile gateway, and the WiFi or cellular connection (green) between the mobile device and the remote cloud server.  }
\label{image:sys_diagram}
\end{figure}

In addition to the {\Sitara}, we develop a mobile application and command and control server interface allowing hundreds of devices to operate in coordination, as shown in Fig.~\ref{image:sys_diagram}. The server provides a convenient web-based GUI for live monitoring and control of connected clients (Fig.~\ref{image:page}) and processing of historical measurement data. This allows monitoring real-time measurements in a distributed, mobile environment or delayed logging and upload for later analysis. Accessible records contain RSSI measurements, IQ samples, location, time and device ID. These capabilities enable passive crowd-sourced measurements using remote control, or to function as a standalone SDR, controlled wirelessly through a user's mobile device. This architecture provides many benefits which allow adaptation for use with additional devices and increased processing and analysis capabilities. For example, abstracted software interfaces and open protocols allow straightforward adaptation for use with other sensor platforms and software-defined radios; the server also provides a convenient platform to run and test various transmitter localization algorithms.

\begin{figure}[ht]
%\raggedright
\includegraphics[width=8.4cm]{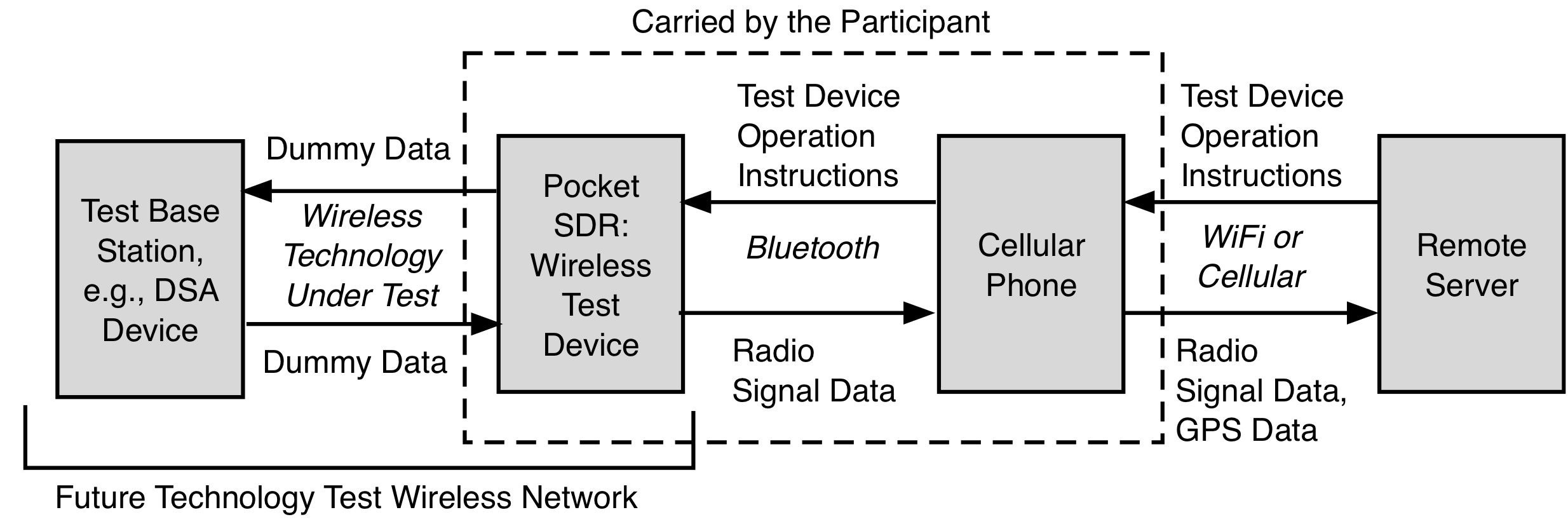}
\caption{{\Sitara} measures signals from other {\Sitara}s or other test devices, and uses the phone and its data connection and GPS to log data and receive commands. }
\label{image:flowchart}
\end{figure}

In this paper we provide a system overview discussing the concept and function of the {\Sitara}. We then share implementation details, design choices and technical specifications. Experimental results from the following usage scenarios illustrate the utility of our SDR: 
\begin{itemize}
\item Transmitter localization using RSSI measurements
\item Crowd-sourced measurements using multiple concurrent participants, suitable for spectrum monitoring or RF propagation modeling
\item Server-side demodulation from IQ sample captures of a 2-FSK transmission
\end{itemize}
We conclude by comparing our solution to existing systems and discussing future areas of research related to our crowd-sourced measurement approach.

%% file: parts/overview.tex
\section{System at a Glance}
The system as presented consists of three primary components depicted in Fig.~\ref{image:sys_diagram}. These are the {\Sitara}, app on the mobile gateway (typically a smartphone), and the remote command and control server. The {\Sitara} produces measurement data which is then sent to the mobile gateway over Bluetooth. The Crowdsourcer application on the mobile gateway either uploads the measurement data to the server over a cellular or WiFi connection or logs the data for upload at a later time. This architecture is designed to accommodate hundreds of simultaneous users. We provide a brief overview of these three components and later continue with implementation details in Section \ref{section:design}.
%, placing a greater emphasis on the {\Sitara} as this is where most of the development effort occurred. 

\subsection{\Sitara}
The {\Sitara} functions as the mobile low-power software-defined radio, responding to commands (Table \ref{table:commands}), performing RF signal capture and spectrum measurements, and processing acquired data for transfer to the mobile gateway and, subsequently, the server. The {\Sitara} features a single circuit board populated with a Texas Instruments CC1200 transceiver acting as the radio and a low-power Nordic nRF52840 SoC as its $\mu$C. The SoC is equipped with a Bluetooth 5 transceiver to maintain high throughput of the raw IQ samples generated by the Texas Instruments CC1200 transceiver~\cite{luong2016rss, Luong:2018:STF:3207947.3207959}. The {\Sitara} relies on a Bluetooth connection for operation.

\subsection {Mobile Gateway - Crowdsourcer}
Local and remote control of the device operate over a Bluetooth link to the mobile gateway. The mobile gateway hosts our Android Crowdsourcer application. This application runs on the mobile gateway, sends commands and receives information from the {\Sitara}. The Crowdsourcer application is designed to capture and log samples from several different RF measurement devices. For the {\Sitara}, the application permits two configurations: standalone manual operation and unattended remote control. In remote control, the server interfaces with the application to issue commands to specific devices with no need for input from the participant. During local operation the mobile owner is able to issue commands and view measurements using the Crowdsourcer application.

\subsection{Command and Control Server}
The Crowdsourcer application communicates with a cloud-based server. All requests are served by a Python Flask~\cite{flask} application through a Gunicorn Web Server Gateway Interface (WSGI)~\cite{gunicorn}. An Apache server is used as a pass-through for security on the front-end. Our application serves the web-based GUI for command and control as well as back-end socket communications among clients. A database stores uploaded measurements in addition to their timestamp, latitude, longitude and unique device ID. Uploaded measurement data may consist of RSSI or complex baseband (IQ) sample data. The server provides a client list to show current connections as well as a page for measurement data.The web-based GUI provides an embedded Google Map object to monitor sensor location and an input terminal for issuing commands. Additionally, the server accommodates scripts to automate commands and test scenarios --- a feature which becomes invaluable for the types of experiments we present in section \ref{section:results}.

%% file: parts/implementation.tex
\section{Design and Implementation}
\label{section:design}
In this section, we examine some of the technical challenges and design decisions during the development of the {\Sitara}, beginning at the hardware component level and then continue with the firmware development. We also briefly discuss the software development associated with the mobile gateway and server applications.

\subsection{Component Choice}

In order to develop a low cost, low power solution we look at transceivers capable of RF digital sample output. After considering many options we tended toward wireless transceivers such as the TI CC1200, Atmel RF-233, AT86RF215,  Atmel AT86RF215IQ, and Silicon Labs EFR32FG. Among these, interface options and operational frequencies lead us to the CC1200 which tunes to frequency bands between 137~MHz and 950~MHz~\cite{cc1200}. The CC1200 is energy efficient and allows raw IQ samples to be exported while still operating over a wide enough frequency range to prove useful.
%mobile sdr capable chipset
%cc1200 - subGhz
%rf233 - 2.4ghz
%atmel rf215iq - sub ghz and 2.4ghz 13bits I/Q 4MHz, need to pair with %FPGA or powerful microprocessor
%EFR32 - subghz and 2.4 SOC, bluetooth 5 advertisment only, low data %rate
The RF network can be configured to match the frequency and bandwidth of operation. In this paper, {\Sitara} uses a 915 MHz balun, which places the optimal operating frequency between 902 to 928~MHz. Future development could add an RF switch or wide-band matching network to improve performance across other bands.

The CC1200 transceiver operates using a SPI interface to read and write data and control registers, respectively. General-purpose IO (GPIO) connections between the CC1200 and SoC $\mu$C allow interrupt-driven functions such as IQ sample acquisition and RF power level triggering. We adapt the register configurations for optimal spectrum monitoring. The CC1200 provides 3 registers (17 bits total) of magnitude and 2 registers (10 bits total) of angle measurements from the output of its coordinate rotation digital computer (CORDIC) algorithm.  

We choose the nRF52840 SoC because it was one of the first available low-power Bluetooth 5 SoCs with a well-supported SDK~\cite{nrf}. Additionally, the ARM Cortex-M4 within the SoC provides a floating point unit (FPU) which is necessary for some SDR applications. Once powered on, the SoC initializes the CC1200 transceiver and executes the program stored in flash. The PPI together with the DMA controller allows many tasks to be automated independent of the $\mu$C. In our application this allows repeated magnitude and angle sample reads from registers on the CC1200 to be loaded directly into RAM allocated as receive buffers within the program. This feature is essential to provide uniform sample capture for DSP applications which would otherwise be subject to interrupt-driven $\mu$C preemption by the Bluetooth stack, resulting in jitter and other non-deterministic timing anomalies. While all sample acquisition and storage is accomplished in hardware, a software interrupt is necessary to switch buffers. 

The RF output from the nRF52840 is connected to a 2.4 GHz 3dBi SMD chip antenna~\cite{bt_antenna}.  While chip antennas are inefficient, the use case is to have a very short Bluetooth link, for example a volunteer might carry both devices in the same handbag, or in two different pockets. Such short links can be reliable even with the antenna loss as we can see from the Bluetooth throughput measurements in section \ref{section:throughput}.
A power management IC (PMIC) regulates voltage, charges and manages the LiPo battery connected through the standard JST connector. {\Sitara} contains a JLink interface to allow programming, terminal logging and debugging. The CC1200's RF chain interfaces with an RF-tuned circuit terminating on a $\mu$.FL connector. For our experiments and the results presented here, we use a Yageo Penta-Band WWAN antenna \cite{antenna}, but other antennas could also be used.
The board, battery, and antenna are designed to fit within a standard 70 by 50 by 20~mm plastic case, which provides mechanical protection while being carried by a volunteer.  The battery is recharged by the volunteer using a standard micro USB cable, likely to be familiar to an Android phone user.

\begin{figure}[ht]
\includegraphics[width=7.4cm]{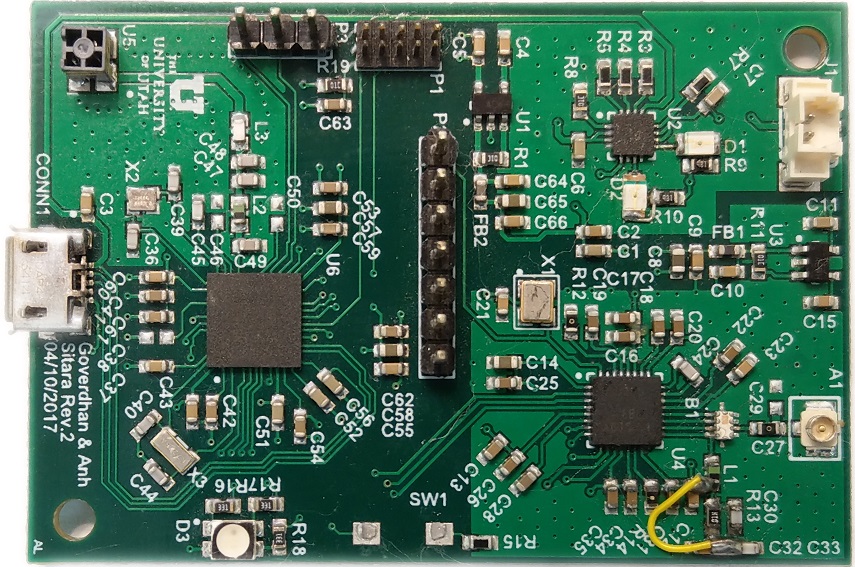}
\caption{{\Sitara} printed circuit board}
\label{image:pcb}
\end{figure}

At the time of writing, the total cost for the bill of materials (BOM) in quantities of 1000 was estimated to be \$38.00 per device. Please refer to our github repository to view the current BOM, source code and design documents: \url{https://github.com/SPAN-UofU}.

%qtn 1k
%PCB 1.8239999999999998 USD
%Parts 33.18495 USD
%Case 1.55 USD
%Antenna 1.14 USD
%Battery 0.5 USD (180mAh) - 0.10 USD (1000mAh)
%total 38

%Note: This reflects pricing (QTY = 1000) at the time of this document %creation. Battery and antenna are not included due to priced depends %on operation requirement. For more up to date pricing, please refer to %the BOM included in the GitHub repo. Also, we do not include one time %fees as it is difference due to the varieties of assembly and %engineering costs based on the contract manufacturer.

\subsection{{\Sitara} Firmware}
We develop the SoC firmware using the nRF5 SDK v13.0.0 from Nordic Semiconductor~\cite{nrf}, compiled using the GNU ARM toolchain v7.2.1~\cite{arm}. The firmware executable code resides in flash on-board the nRF52840 SoC. The SoftDevice, a pre-compiled protocol stack, is also stored in flash and loaded into RAM at run-time. An event-driven API allows the firmware to interface with the SoftDevice to access Bluetooth functions.

\begin{figure}[ht]
\includegraphics[width=8.4cm]{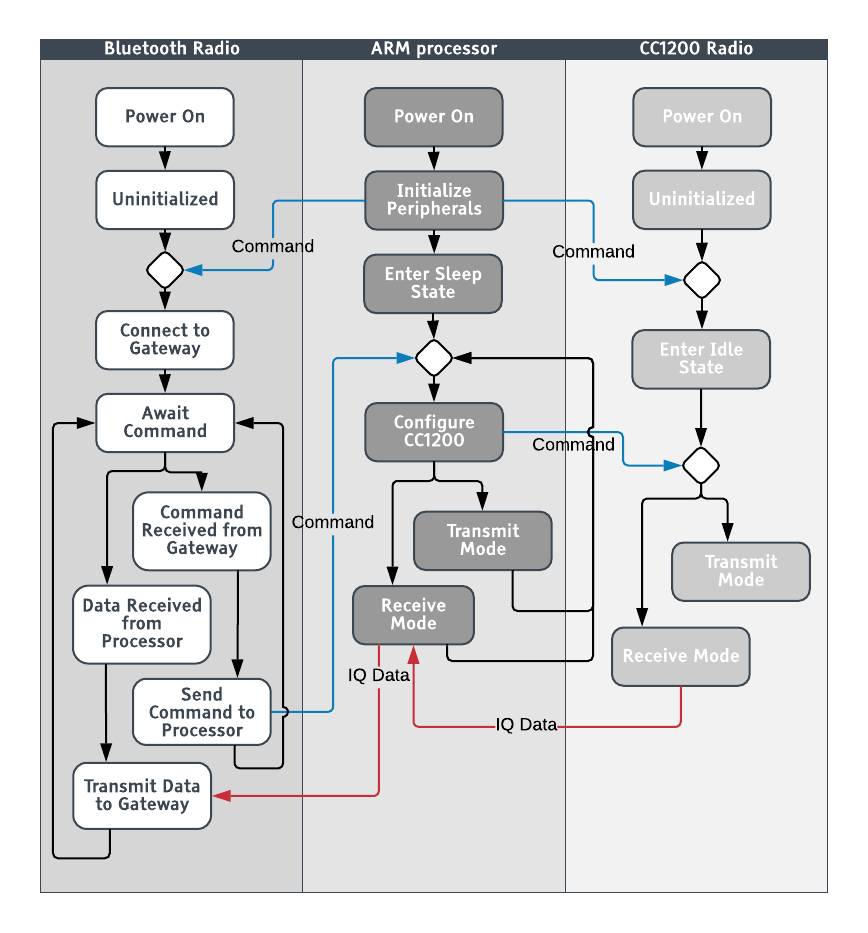}
\caption{Simplified state diagram for {\Sitara}}
\label{image:state}
\end{figure}

Once powered on, the $\mu$C initializes and configures the external CC1200 transceiver, on-chip Bluetooth radio and other peripherals then enters a sleep state while awaiting commands. As we mention, minimizing power consumption is a key design driver, so minimizing the time that components are powered on and active is a recurring theme. This allows us to achieve an 80\% power reduction for most applications. As a result, most functions and all commands are event-driven. Commands are received on incoming Bluetooth packets, triggering an event that wakes the $\mu$C and enters an event handler which parses the message and extracts a valid command. These commands are encoded as ASCII characters and correspond to the functions listed in Table \ref{table:commands}; a state diagram is shown in Fig.~\ref{image:state}.

\begin{table}[ht]
\captionsetup{position=below, skip=10pt}
\begin{tabular}{ |p{1.9cm}  p{5.9cm}|}
\hline
 Command &  Description\\
\hline \hline
ACCESS REGISTER & Allow access to any (or all) registers in the CC1200\\ \hline
SAMPLE CAPTURE & Begin continuous or carrier-triggered sampling of magnitude, angle or RSSI and upload values to gateway via Bluetooth\\ \hline
TUNE FREQ & Tune the radio to a user-provided frequency with 1 kHz resolution\\ \hline
CHANGE PHY & Increase the Bluetooth data rate to 2~Mbps for compatible devices\\ \hline
TRANSMIT & Transmit provided text string over the air using a supported modulation scheme or continuously transmit a single RF tone\\ \hline
RECEIVE & Receive incoming transmissions using a supported modulation scheme and send the results across the Bluetooth link\\ \hline
LOCK FREQ & Attempts to phase lock to the strongest signal near a selected frequency\\ \hline %initiates a frequency lock function which first scans and tunes to the frequency of highest-measured RSSI, then performs successive phase measurements and tunes the frequency until either the phase accumulation over the sample period converges to a minimum or the algorithm times out\\
STOP & Halt any current measurement or transmit operation and return the radio to a low-power idle state\\ \hline
RESET & Reset the {\Sitara}\\ 
\hline
\end{tabular}
\caption{List of {\Sitara} commands}
\label{table:commands}
\end{table}

Most commands  perform a single function then return the $\mu$C to a sleep state. The SAMPLE CAPTURE command enters a loop in which data is continuously acquired and sent via the Bluetooth interface. Because continuous sample capture is an important aspect of our design we will discuss its operation in more detail. The SAMPLE CAPTURE command allows additional parameters to determine how and which data will be captured and uploaded. These parameters include MAGNITUDE, RSSI-DIRECT, RSSI-CALCULATED, PHASE and TRIGGERED. The differences between these parameters are presented in Table \ref{table:capture} below. 

\begin{table}[ht]
\captionsetup{position=below, skip=10pt}
\begin{tabular}{ p{2.1cm} | p{5.9cm}}
Parameter & Description\\
\hline \hline
\small MAGNITUDE & continuously captures both magnitude and angle (I/Q) data\\
\small RSSI-DIRECT & continuously captures RSSI as reported by the CC1200 with 8-bit resolution\\
\small CALCULATED & continuously captures magnitude and calculates RSSI with 17-bit resolution\\
\small PHASE & continuously captures angle measurements with 10-bit resolution\\
\small TRIGGERED & continuously captures data while RSSI is above a threshold\\
\hline
\end{tabular}
\caption{{\Sitara} sample capture parameters}
\label{table:capture}
\end{table}

Sample capture utilizes the Programmable Peripheral Interconnect (PPI), which permits on-chip peripherals to interact through task-event relationships, independent of the CPU. We configure an interrupt event associated with the magnitude-valid output signal from the CC1200 to trigger a burst-read SPI transaction task which reads the registers containing the sample data. The magnitude-valid signal asserts when a new IQ sample is ready on the CC1200 and occurs at a set rate dependent on the configured receiver filter bandwidth. The sample data automatically load into a pre-allocated receive buffer using the PPI and DMA. The completion of the SPI transaction triggers another event which increments a counter using the PPI. Once this counter reaches a predetermined count, another task begins which temporarily disables the capture function and triggers a software interrupt event. The purpose of this software interrupt is to assign the pointer to the receive buffer to a new memory address so sample capture can resume while the previously-filled buffer is processed and sent over the Bluetooth interface. Unfortunately, the nRF52840 SoC does not provide a method of switching receive buffers for the SPI-DMA transaction without $\mu$C intervention; however, the time required to service the interrupt is approximately 22 $\mu s$. Therefore, sampling rates with a period equal to or greater than 22 $\mu s$ (45 kS/s) may be delayed by at most 1 sample period. An additional factor that complicates this process and limits the maximum sampling rate over SPI is the absence of a hardware-enabled control function which allows consecutive burst reads from multiple registers (although this is available for repeated reads from a single register). This negatively impacts performance in two ways: 
\begin{itemize}
\item Each sample read must include two command bytes for SPI burst read which are added to the total length of the transaction, thus increasing each transfer duration. 
\item During a SPI transfer, a byte is received during each clock cycle, thus the receive buffer will always contain the two status bytes received during the clock cycles which the two command bytes are sent, in addition to the bytes containing the actual data.
\end{itemize}
This results in inefficient use of memory and a receive buffer containing status bytes interleaved with sample data requiring extra processing steps to extract samples. Nevertheless, this does not impact performance because the Bluetooth connection ultimately limits throughput as we discuss in Section \ref{section:throughput}.

We define three rotating buffers to store received samples. This allows SPI transfer using DMA to occur in one buffer while simultaneously processing another in preparation for Bluetooth transmission. The third buffer stores the processed data awaiting awaiting Bluetooth transmission. Prior to transmission over Bluetooth the samples are compressed to increase transport efficiency and improve the data rate. During sample capture, up to 5 bytes are read from the magnitude/angle registers but the values are represented by 17 and 10 bits, respectively. Because the Bluetooth serial service operates on bytes and supports flow control and acknowledgement, we compress this data into 3 bytes as follows.  We observed through initial experimentation that the 3 least significant bits of magnitude are mostly measurement noise, thus we remove them from the magnitude and keep the full 10 bits of angle, for a total of 24 bits. The number of captured samples per array and array size were chosen to fill a maximum-length Bluetooth packet data payload of 244 bytes. During IQ sampling 81, 5-byte samples are captured, sent over the SPI interface and then compressed to 3-byte samples to fit into the 244 byte Bluetooth packet payload. This process is shown in Fig.~\ref{image:buffers} below.

\begin{figure}[ht]
\includegraphics[width=8.4cm]{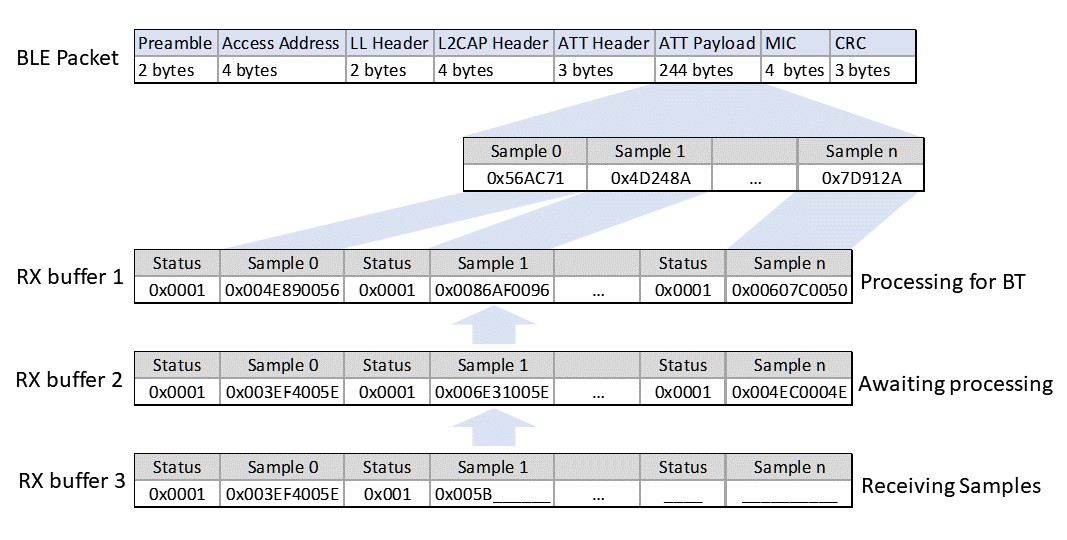}
\caption{{\Sitara} buffer system designed to maximize throughput}
\label{image:buffers}
\end{figure}

Bluetooth transfer of packets is not real-time, and due to packet collisions and errors, MAC delays and retransmissions, any finite buffer can experience an overflow. This is handled by pausing capture acquisition and discarding samples while the Bluetooth interface catches up. This is problematic for RF sample capture because it can break continuity and introduce timing offsets. In order to compensate for these errors, we maintain a 16~MHz counter during acquisition which is activated while sample capture is paused and then reports the elapsed pause time once sample capture resumes. This information is then also sent over Bluetooth so the end-user is able to accurately reconstruct and preserve timing of sample captures.

\subsection{Crowdsourcer and Server Software}
The choice of server architecture and software frameworks were driven primarily by convenience, ease of use and adaptability rather than resource optimization as we see for the {\Sitara}. The server application uses the Python Flask micro web framework on an Apache front-end paired with Gunicorn, a Python WSGI server. The control functions are accessed from an interactive Javascript-based GUI in a web browser. Socket.IO libraries for Python, Javascript and Java provide low-latency, standardized event-based communication between different the server, Web GUI and the device gateways, respectively.

\begin{figure}[ht]
\includegraphics[width=7.4cm]{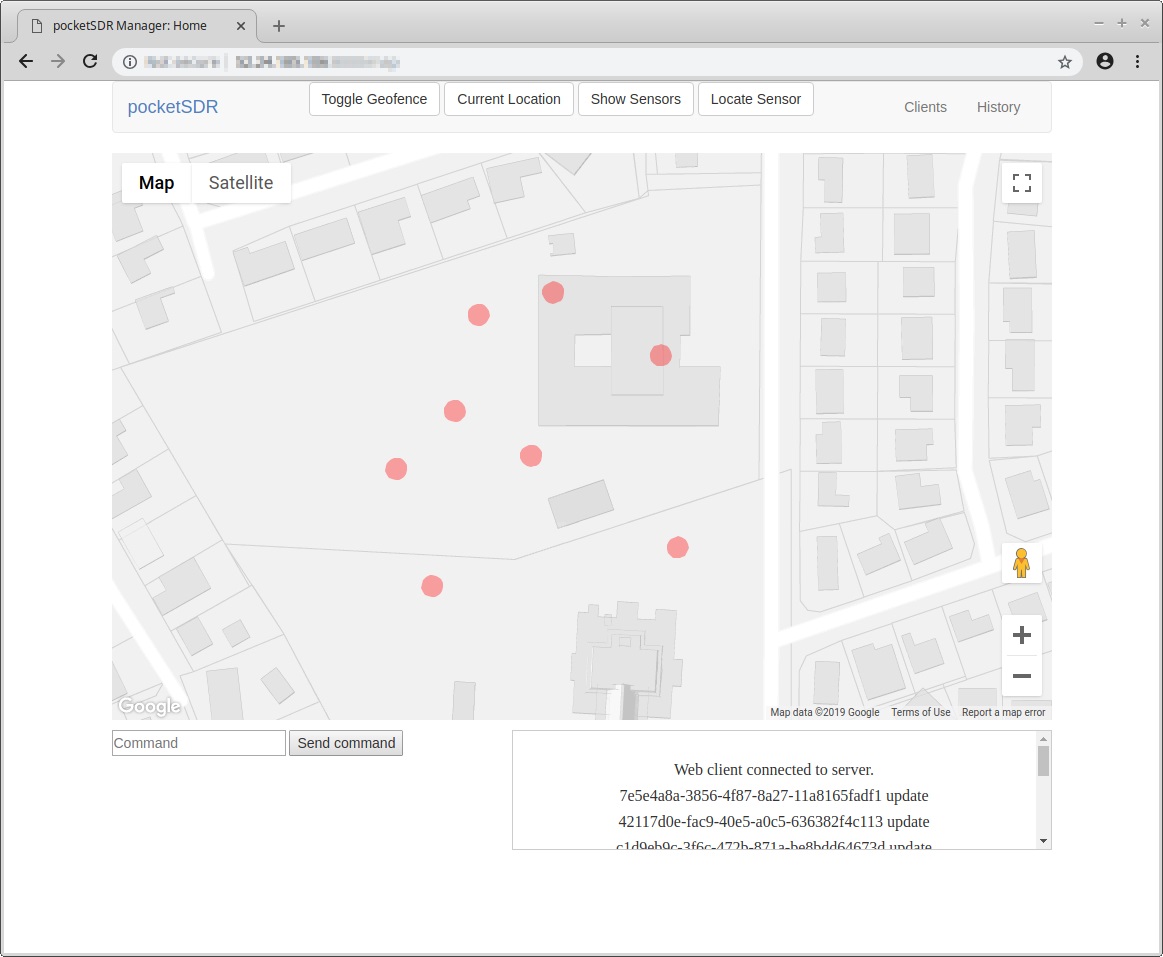}
\caption{Server homepage showing locations of active sensors}
\label{image:page}
\end{figure}

The user issues commands from the web client which are sent to the server and relayed to the appropriate client device gateways. The web client, server, and gateway each activate a set of event listeners which filter relevant messages. Messages containing measurement results in response to commands, such as RSSI and IQ data, are stored in the server's database. The web interface provides a Google Maps overlay which can display real time client location and associated measurement data as shown in Fig~\ref{image:page}. In addition to real-time monitoring of sensors, the web client also provides convenient tools for filtering and displaying subsets of measurements according to a number of parameters such as time, frequency, RSSI threshold and location. Some of these capabilities will be demonstrated in section \ref{section:results}.

Messages to and from the server are formatted as JSON data relevant to the context. Using the Crowdsourcer application installed on the mobile gateway, the user initiates a connection to the server which persists until manually terminated at either end. The server can issue the following commands:
\begin{itemize}
    \item Report RSSI - commands the {\Sitara} to report measured RSSI values at a user-configured rate
    \item Triggered Capture - commands the {\Sitara} to report measurements once an RF threshold is achieved
    \item Continuous Capture - commands the {\Sitara} to continuously upload IQ data
    \item Clock - synchronizes the {\Sitara} clock with GPS time obtained by the mobile gateway
    \item Stop - disables any capture mode and returns {\Sitara} to idle state
    \item Transmit - commands the {\Sitara} to transmit a CW signal or modulated message
    \item Upload - upload captured data beginning at user-specified time point
    \item Debug mode - sends arbitrary commands directly to the {\Sitara}
\end{itemize}

The Server application was designed to provide a high-level abstraction for commands, leaving the implementation details and low-level commands to the client device gateways. For example, a server RSSI command emits a single message containing several parameters such as frequency, bandwidth, reporting interval and report type. The device gateway receives this message and issues multiple commands as appropriate, such as frequency tune and RSSI capture, over Bluetooth to individual {\Sitara} devices. A {\Sitara} in turn will interpret each of these commands and perform additional steps to accomplish this task. Combining and abstracting commands at the top level in this way reduces latency, preserves bandwidth and spares server resources by minimizing the number of message exchanges for a given task. This abstraction also provides a modular approach making the sensor implementation-agnostic to the server.

While developing this architecture, we considered implementing an existing standard such as IEEE 802.22.3 Spectrum Characterization and Occupancy Sensing (SCOS) Sensor~\cite{scos}, an extension of the SigMF specification~\cite{sigmf}. We find that such specifications provide a level a complexity beyond our immediate needs. Specifically, SCOS implements a restful API requiring each sensor to host a web server; we instead use the Socket.IO protocol to maintain persistent connections necessary for managing mobile devices. Notwithstanding the differences, our implementation could be adapted to comply with the SCOS standard as follows: Rather than implementing a RESTful API on each sensor, a gateway could be installed on the server which is compatible with the SCOS standard and emulates the API for each device. As requests are received at this server, the SCOS gateway would forward the appropriate commands to the server application. This would provide all the benefits of compatibility with the SCOS standard without a system redesign.

%% file: parts/evaluation.tex
\section{Evaluating Our Solution}
We now present measurements characterizing the performance of our system under varying conditions. A large enough sample size of {\Sitara} devices was not available to thoroughly characterize the statistical variability for mass production/manufacturing purposes, so instead we present these results as a reasonable expectation of performance.

\subsection{Data Throughput}
\label{section:throughput}
Three data paths potentially limit the real-time throughput of the {\Sitara} system. The Bluetooth data rate, the SPI interface from the CC1200 radio to the $\mu$C, and the on-board $\mu$C itself. Depending on the use case, any of these could become a bottleneck. In our application, the maximum data transfer across the SPI bus exceeds the data rate of the Bluetooth link.  

The CC1200 specifications limit the minimum SPI clock rate to 7.7 MHz for extended register reads, which include the registers of interest. In practice we have successfully achieved an 8 MHz clock rate. Two common modes of operation read from the CC1200 either three magnitude and two angle registers (8-bits each)  or only the two angle registers. The maximum achievable sampling rates for these two modes were 64 kS/s and 104 kS/s, respectively. This was determined experimentally by gradually increasing the sampling rate while measuring the actual transfer rate on an oscilloscope until data transfer ceased. These sampling rates were achieved using the programmable peripheral interconnect (PPI) of the nRF chip which allows a burst (consecutive) read of multiple registers --- up to five in this case. This allows a faster and more consistent (in terms of timing) sample transfer than would be achievable with $\mu$C software interrupts.

\begin{figure}[ht]
    \centering\includegraphics[width=7.4cm]{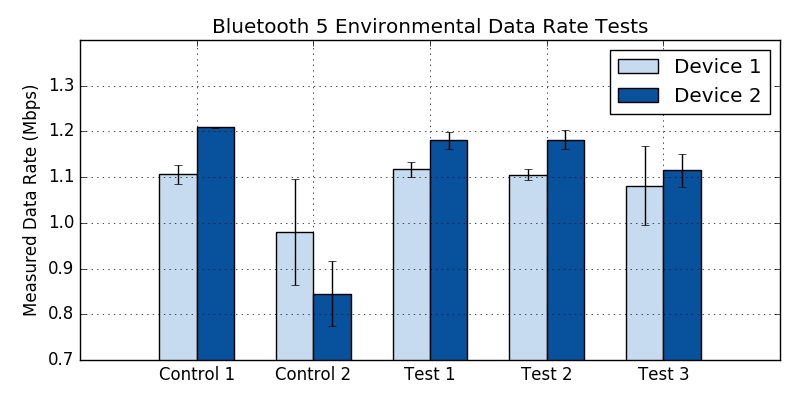}
    %\caption{Comparison of maximum {\Sitara} energy consumption by operational state}
    \hfill

    \centering\includegraphics[width=7.4cm]{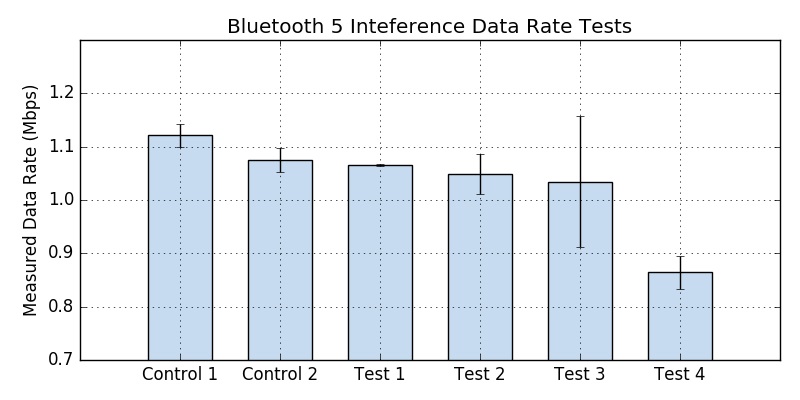}
    \caption{Measured Bluetooth 5 data rate between {\Sitara} and two mobile devices under varying test scenarios (top). Measured Bluetooth 5 data rate in the presence of varying numbers of interferers, where test number refers to the number of interferers (bottom). All error bars represent standard deviations.}
    \label{image:bluetooth}
\end{figure}

The Bluetooth 5 standard defines a maximum transfer rate of 2~Mbps~\cite{blue}. By utilizing this LE 2M PHY option, packet length extension and configuring the Bluetooth Maximum Transmission Unit (MTU) to match the packet length, we achieved highest throughput.  Initial throughput testing using special firmware achieved a peak effective data rate of approximately 1.3 Mbps between a {\Sitara} device and Bluetooth 5 capable mobile phone. In typical usage scenarios we observe an average throughput greater than 1 Mbps. This determination results from a series of tests under varying environmental conditions using two different mobile phones, denoted \emph{device 1} and \emph{device 2} in  Fig.~\ref{image:bluetooth}. For each of these measurements, 244KB of data was transmitted over the Bluetooth link while the {\Sitara} recorded the transmit time on its system clock. The 244KB transmission was then repeated at least 10 times for each test. For the environmental test (Fig.~\ref{image:bluetooth}, top) we used two controls, the first places a mobile phone within 15 cm of the {\Sitara} in an environment without in-band WiFi or Bluetooth activity as observed on a spectrum analyzer. The second control repeats this test, but separates the paired devices by 12 meters. We then perform additional measurements as follows: with the participant keeping the {\Sitara} in a pocket while holding the phone in hand (Test 1), with the {\Sitara} inside a backpack while the participant holds the phone in hand (Test 2), and finally, in a variety of different indoor and outdoor environments with the {\Sitara} in the participant's pocket (Test 3). In addition to the environmental tests, we also conduct interference tests (Fig.~\ref{image:bluetooth}, bottom). For these experiments, up to four interfering devices transmit over Bluetooth at their maximum data rate while the unit under test performs measurements. For these tests, the paired devices are separated by one meter. The controls consist of the same test, with no interferers (Control 1) and four interferers (Control 2); though in the case of the latter, the separation is reduced to 15 cm while other conditions remain the same. Test 3 of the interference measurements notably exhibits higher deviation than others. We speculate that this is due to characteristics of the Bluetooth protocol at marginal capacity or differences in hardware among devices. Further analysis of the Bluetooth performance and characterization is beyond this scope of this work as these tests are intended only to approximate worst-case conditions expected in real-world scenarios. 

In our application, minimal processing was required by the onboard $\mu$C during sample acquisition and was not found to impact throughput, therefore no attempt was made to evaluate load or processor utilization on the {\Sitara}. If additional signal processing were to be carried out on the {\Sitara}, then this may require further investigation.

\subsection{Power Characteristics}
The {\Sitara}'s power consumption ranges from approximately 18 mW to 180 mW depending on operational mode. Power usage in typical scenarios is shown in Fig.~\ref{image:power} below. In the {\Sitara} idle state, the CC1200 radio is set to a low powered state, maintaining power only to the crystal oscillator and digital core; when no other commands are present the $\mu$C in the nrRF52840 chip receives the wait-for-event (WFE) command which likewise powers down nonessential modules. The $\mu$C will periodically wake up to handle events necessary to maintain a Bluetooth connection. During RX and TX states, the CC1200 remains powered on along with necessary RF, clock and interface peripherals; the $\mu$C is more heavily utilized in these states, but no quantitative analysis was performed to determine the duty cycle of the active versus inactive/WFE state. It is worth noting that our implementation of the continuous triggered measurement mode continuously captures RF samples but will only retain and send the data across the Bluetooth link if the RSSI threshold is reached. This is not particularly energy-efficient but was implemented this way to maintain a brief history of samples before the trigger event. A much more energy-efficient implementation is possible which would allow the ARM core to sleep while awaiting a threshold detection trigger event from the CC1200, at which point sample acquisition would begin. The trade-off is that this would introduce a delay between the trigger event and sample capture that could be as long as a millisecond. For spectrum measurements this may suffice, but for IQ sample measurements, a signal of interest could be missed during this delay.  

\begin{figure}[ht]
\centering\includegraphics[width=7.4cm]{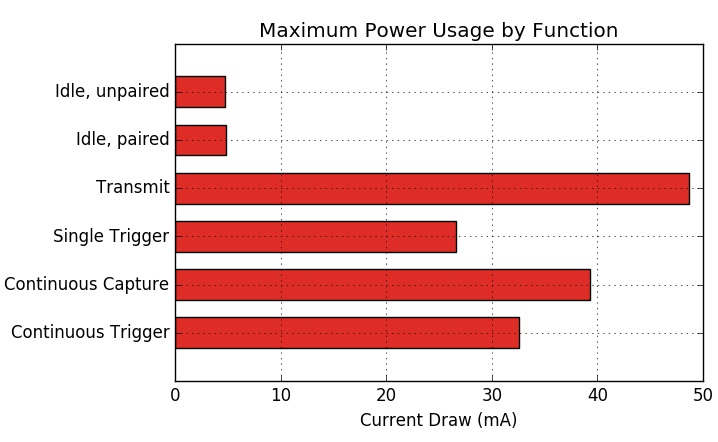}
\caption{Comparison of maximum {\Sitara} energy consumption across a range of operational states.}
\label{image:power}
\end{figure}

% \begin{table}[ht]
% %\setlength{tabcolsep}{0.5em}
% \begin{tabular}{p{0.75cm} | p{4.4cm} | c}
% State & Function & Current (mA)\\
% \hline \hline
% Idle & not connected & 4.6\\
% Idle & Bluetooth connected	& 4.7\\
% RX & triggered measurement	& 26.5\\
% RX & mag/angle measurement & 39.1\\
% TX & max power continuous wave &	48.6\\
% \hline
% \end{tabular}
% \caption{Current consumption measurements}
% \label{table:power}
% \end{table}

We initially tested battery life with a 3.7 V 180 mAh LIPO battery, allowing the {\Sitara} to operate for several hours in any state. Repeated tests demonstrate that at full charge, the {\Sitara} maintains a consistent Bluetooth connection until loss of power after 37 hours; this is consistent with the estimated battery life based on the measured idle power consumption. After testing we chose to replace the 180 mAh battery with a 850 mAh battery to extend battery life up to a week with a commensurate improvement in high duty-cycle operation. For comparison, the USRP E312 SDR uses a 3200 mAh battery to achieve 5.5 hours at idle~\cite{ettus}.

\subsection{Timing and Synchronization}
The CC1200 contains a phase-locked loop (PLL) which allows timing recovery in hardware for available modulation schemes. In other applications, our firmware performs a phase-lock function to effectively tune to a signal of interest. While the {\Sitara} proved effective for our scenarios, there remains a temperature dependency in the oscillator which could present a challenge for long term clock stability in some applications. Use cases requiring higher clock stability and accuracy such as Doppler, time-of-arrival, or synchronous RF network architectures may not be possible or may required frequent re-synchronize of devices. A future board revision will incorporate a more accurate temperature compensated crystal oscillator to mitigate these effects.

For system timing and timestamp information relating to signal measurement, the {\Sitara} maintains a system clock. The {\Sitara} can synchronize its clock by receiving a clock command containing current GPS time from the mobile gateway over Bluetooth. Multiple such clock commands are sent until the mobile gateway observes a minimum round-trip time between the sent time and command acknowledgement --- a technique not very different from many network time protocols. This provides a simple method of synchronizing devices with GPS time with an error on the order of milliseconds. In order to achieve a more accurate synchronization, {\Sitara} devices can perform a triggered RSSI measurement which will upload a timestamp associated with the trigger event to the server. From a collection of triggered RSSI measurements on the server for a single transmit event, relative clock offsets between different devices can then be easily computed. If an application requires a more accurate node synchronization, then a more sophisticated approach would be appropriate \cite{synch, chorus}. 

\subsection{Server Performance}
Although some performance compromises are necessary to meet our hardware design goals for the {\Sitara}, server-side resources present few practical constraints as they are easily re-configurable at run-time. By hosting our server on a cloud-computing service, our application can quickly scale to meet demand. As a base configuration we reserve a host with 1 vCPU and 0.5 GB of RAM. We find this adequate to display and manage 250 simultaneous clients each reporting one RSSI measurement per second. Additional clients reporting RSSI, or multiple clients uploading IQ samples may require a more capable server if the live reporting mode is desired. As our system is expected to handle dropouts and latency associated with mobile data links, we place no hard timing requirements on server resources and assume best-effort. 

%% file: parts/results.tex
\section{Case Studies}
\label{section:results}
In this section, we present experimental results demonstrating how our system is used in two different applications. We specifically choose these applications to demonstrate the versatility and key features of our system namely: the portability of the {\Sitara} for crowd-sourced measurement and its utility as an SDR. We also explain how these demonstrations can extend to much more complicated experiments. In our two examples we use the terms \emph{local} and \emph{remote}, respectively, to denote a user who manually performs measurements using the mobile gateway at the location of measurement and a user who issues a command from the server to perform measurements at a remote location where a {\Sitara} is present. Additionally, in a crowd-sourced environment, we say a participant can be \emph{active} or \emph{passive}, respectively, meaning the participant has some information or direction from the user conducting the experiment or the participant has no knowledge about or communication with the user conducting the experiment. For example, if a user conducting an experiment instructs participants to sweep location $X$ to find a transmitter, then these would be considered active participants. 

Note that for experiments involving volunteers, to ensure consent and handling of potentially sensitive user data is adequately addressed, we maintain an institutionally-approved IRB.

\subsection{Spectrum Sensing}
One of our primary design goals for the {\Sitara} is to offer a convenient platform for crowd-sourced spectrum sensing. As such, we demonstrate the versatility of our system in this capacity in real-world settings. \\

\subsubsection{Transmitter Localization: Single {\Sitara}}
In this scenario, we deploy one {\Sitara} acting as a receiver which is controlled locally using an open-source third-party mobile application: \textit{nRF UART v2.0}. The user walks around the test site, issues commands from the mobile gateway to capture RSSI measurements which are later used to estimate the location of a ``rogue" transmitter. 

A USRP B210 radio serves as a transmitter, broadcasting a 910 MHz continuous wave signal at 20dBm as measured at the antenna interface. Considering the typical output power of consumer wireless devices and the propagation characteristics within this frequency band, 20dBm would be a conservative value bordering on worst case for the purposes of detecting a disruptive transmitter; a malicious actor, intent on disruption, would likely employ a more powerful transmitter to degrade performance of devices over a wider area. The transmitter broadcasts through a 3dBi vertical omni-directional antenna. 
%\begin{figure}[ht]
%\includegraphics[width=8cm]{parts/experiment_setup.jpg}
%\caption{A USRP B210, broadcasting a 910 MHz continuous wave signal was used as the "rogue" %transmitter for this experiment}
%\label{image:setup}
%\end{figure}

\begin{figure} [ht]
    \centering
%    \begin{subfigure}
        \centering\includegraphics[width=4.1cm]{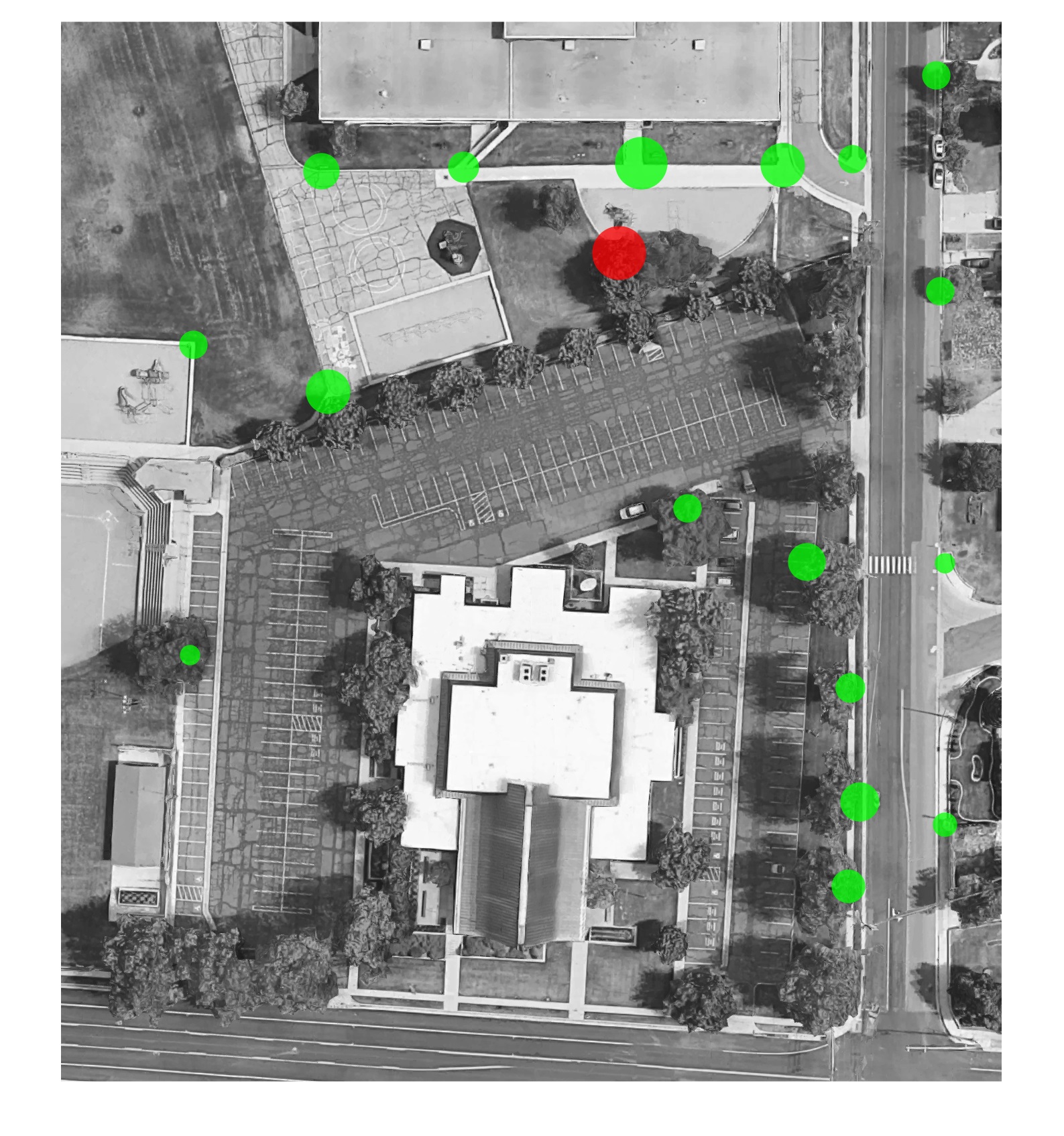}
        %\caption{The test area with transmitter location shown in red, and measurement locations shown in green}
%        \label{image:site}
%    \end{subfigure}
    \hfill
%    \begin{subfigure} 
        \centering\includegraphics[width=4.1cm]{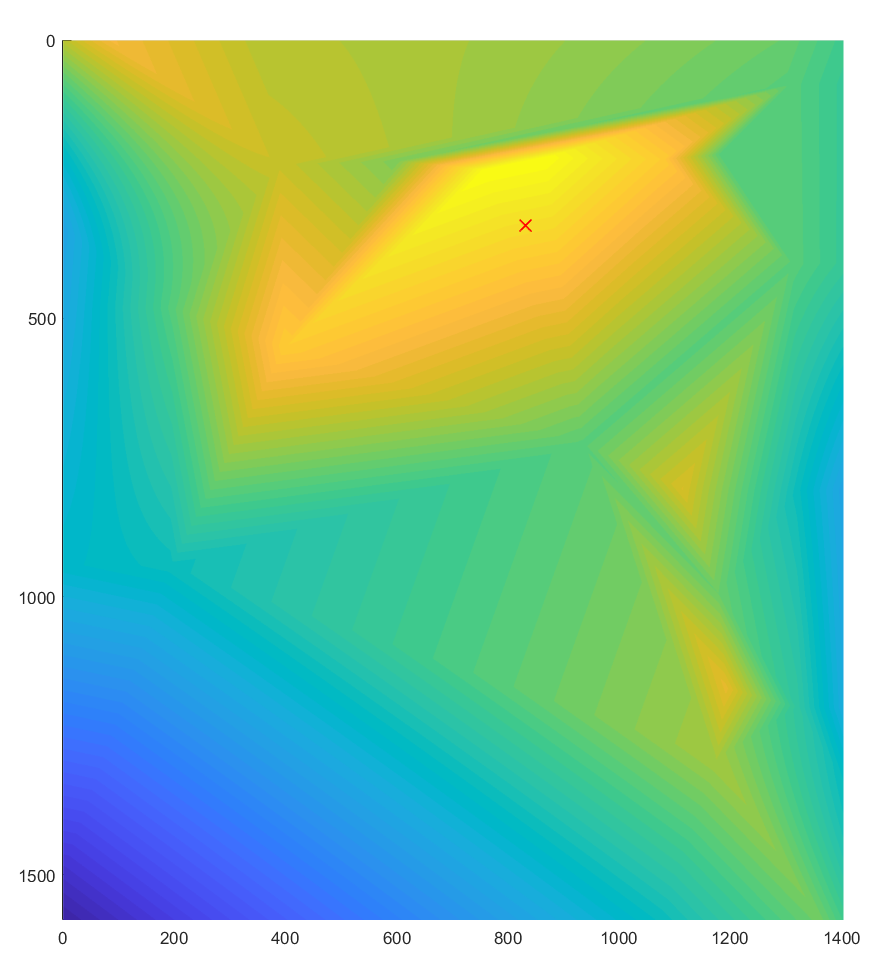}
        %\caption{Transmitter location estimation; the red $x$ denotes the true location}
%        \label{image:interp}
%    \end{subfigure}
    \caption{(left) Outdoor test area with transmitter location (\textcolor{red}{$\bullet$}) and measurement locations (\textcolor{green}{$\bullet$}), with radius proportional to the measured RSSI value. (right) Transmitter location estimation map and  true location (\textcolor{red}{$\times$}).} \label{image:site_interp}
\end{figure}

Multiple RSSI measurements are taken to estimate the location of a receiver. The measurement points are chosen arbitrarily along pedestrian-accessible paths. The measurement points used for localization are indicated by green circles overlaid onto the site map in Fig.~\ref{image:site_interp} (Left). The radius of the green circles are proportional to the measured RSSI value at each of these points. The red circle denotes the transmitter location. The test site covers a roughly 28,000~$m^2$ area, with the farthest discernible measurement captured at a distance approximately 120~m from the transmitter. In order to ensure that the intended signal is measured, each sample measurement is recorded after issuing a LOCK FREQ (See Table \ref{table:commands}) command to the {\Sitara} at 910 MHz. Any measurements that do not converge to a phase lock, fall below the noise floor, or fall outside the span of 10kHz centered at 910 MHz are discarded. The user is able to observe the frequency scan and lock in real time.

Once the data is captured, the coordinates of the measurement points are associated with individual pixels of a captured overhead image of the test site retrieved from Google Maps. To make localization more challenging, we choose to discard the highest three RSSI measurements. The remaining points are used to produce Fig.~\ref{image:site_interp} (left), then as inputs to a Matlab interpolant object from which Fig.~\ref{image:site_interp} (right) is generated. In this case, linear interpolation is used. We can see that this small data set performs reasonably well in locating the transmitter. The true location is approximately 9 m away from the best estimate as indicated by the peak of the yellow region in the figure. Clearly with more measurements, either from an increased sampling rate or multiple simultaneous devices, we would see this accuracy improve. %Additionally, using a different antenna with a more uniform pattern could reduce error in location estimation; depending on the orientation of the antenna we used during this test, the non-uniform pattern can result in variations as high as 10dB in the worst-case. 

Here, we deploy a single {\Sitara} with a local, active participant to capture RSSI measurements at different points. Local {\Sitara} operation may be convenient for directed experiments which may not accommodate multiple passive users in a crowd-sourced scenario. A more typical use case will involve multiple participants--active or passive--and rely on automated, server-initiated measurement commands. We demonstrate these capabilities next. \\

\subsubsection{Crowd-sourcing: Multiple {\Sitara}s}
\begin{figure} [ht]
    \centering\includegraphics[width=8.4cm]{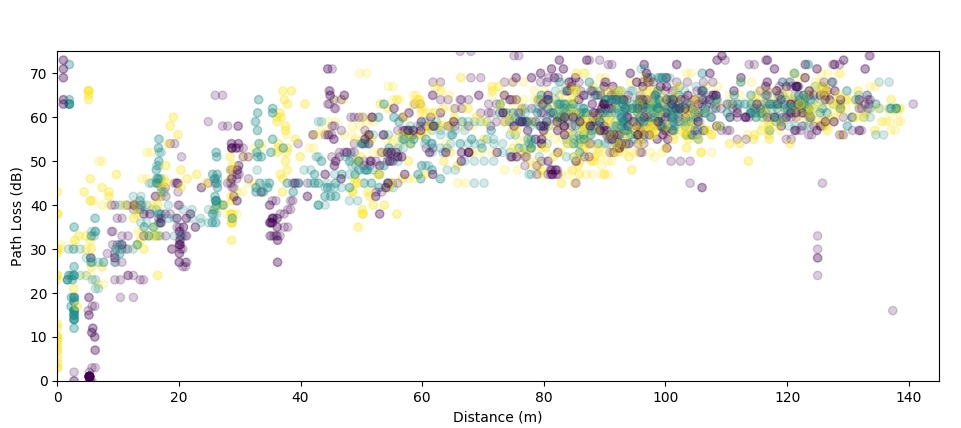}
    \caption{A snapshot in time showing path loss using RSSI measurements from multiple {\Sitara}s, distinguished by color, as a function of distance from a transmitter. Using {\Sitara}, such data sets are easily generated and can be used for propagation models.}
    \label{image:path_loss}

\end{figure}

\begin{figure} [ht]
%    \begin{subfigure}
        \includegraphics[width=4.9cm]{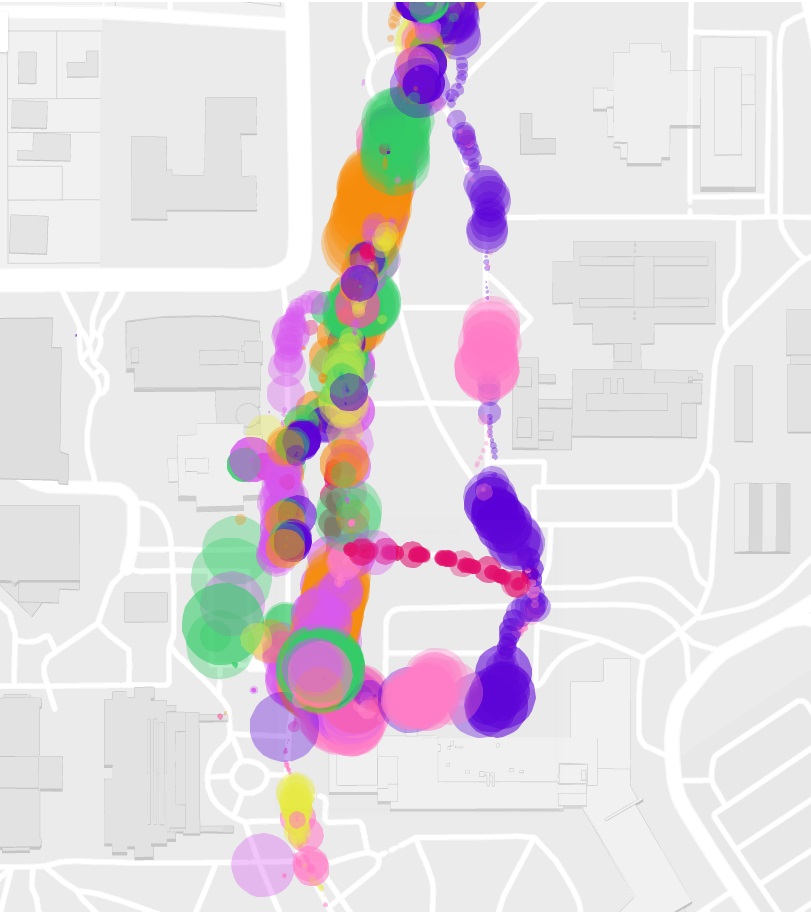}
        \label{image:multiple_devices}
%    \end{subfigure}
    \hfill
%    \begin{subfigure}
        \includegraphics[width=3.1cm]{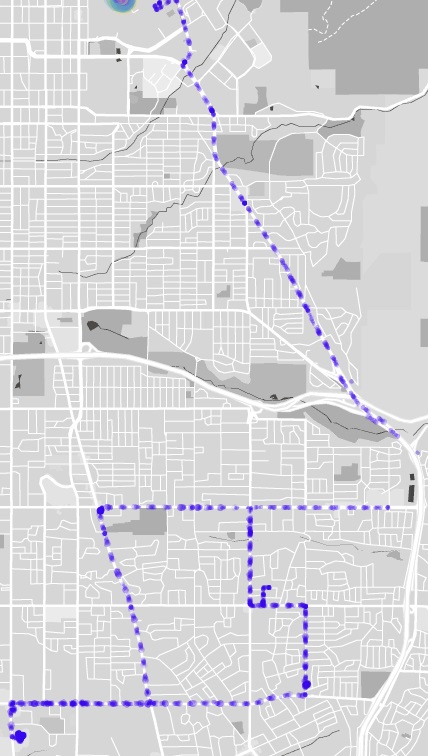}
        \label{image:commute_measurements}
%    \end{subfigure}
    \caption{Server-generated overlays of remote RSSI measurements from twelve devices carried by participants over a series of experiments, where each color represents a unique device and each point radius is scaled relative to RSSI value (left) and server generated overlay of RSSI measurements from one device over a wide-area (right).}
    \label{image:crowd-source}
\end{figure}

We deploy up to twelve {\Sitara}s among passive participants in a series of experiments. Again, by passive we mean the participants are not directed but walk around freely while the nodes are operated remotely from the server, requiring no interaction from participants. By automating different test scenarios using server-side scripts, we are able to rapidly acquire large volumes of data. Fig.~\ref{image:path_loss} depicts path loss in a suburban environment calculated from RSSI measurements and GPS coordinates of multiple devices obtained using a round-robin transmit scheme, in which individual nodes take turns operating as a transmitter while others measure RSSI. This is a fast and efficient method to generate data for inputs into complex propagation models that may otherwise be difficult to obtain \cite{propagation}. Fig.~\ref{image:crowd-source} (left) presents a server-generated overlay of RSSI measurements obtained from twelve {\Sitara}s scanning a range of frequencies over a user-specified duration. Here, each point radius scales relative to RSSI and each color, again, represents a unique device. These measurements could be used for transmitter localization as we demonstrated previously, or as inputs to environmental RF propagation models.   %Our first test uses one {\Sitara} as stationary transmitter with others as receivers. Fig.~\ref{image:crowd-source} (left) shows measurements from three participants as red, blue and green circles with radius proportional to its measured RSSI value. The location of the transmitter is not indicated, but can be estimated from the figure --- roughly centered within the building. This shows that the {\Sitara} can be used indoors for measurements even though the lack of GPS signal causes the mobile gateway to rely on other, less accurate location services such as WiFi or the cellular network. This scenario also demonstrates Crowdsourcer's ability to continue operation as network conditions change. For the red measurements, network service was interrupted as the participant moved into, and later out of, the building and experienced hand-off between WiFi and cellular.

Fig.~\ref{image:crowd-source} (right) shows a server-generated overlay from prior measurements stored in the database using an automated command script. This image is generated by querying the server for measurements (blue dots) from a specific device between two time points. These measurements are not intended to locate a known transmitter, but instead demonstrate the passive crowd-sourcing capability. Regardless, if our object was to determine locations of possible sensors, we could request another overlay including multiple devices and an RSSI threshold. Along one path in the figure, we also see a drop in measurements, this likely indicates a loss of mobile data service in that particular area. In cases where cellular service is less reliable, or where large amounts of data are required, it may be more appropriate to use the logging report mode; during live measurement mode, the {\Sitara} makes a best effort attempt to send measurement data back to the server, but in the absence of a network connection, no data is available. If logging mode is enabled, the {\Sitara} will continuously record measurements, log them to a file on the mobile device and later upload the data upon request from the server. 

% \begin{figure} [ht]
%     \centering\includegraphics[width=8.4cm]{parts/rssi_over_time_RRc.png}
%     \caption{A snapshot in time showing RSSI measurements from multiple {\Sitara}s, distinguished by color, carried by participants during a round-robin transmit scheme. Such measurements, along with their location data, would be suitable for RF propagation modeling.}
%     \label{image:rssi_over_time}

%\end{figure}

% \begin{figure} [ht]
%     \centering\includegraphics[width=8.4cm]{parts/RSSI_combined.png}
%     \label{image:rssi_combined}
%     \caption{RSSI measurements from 12 devices carried by participants over a series of experiments, where each color represents a unique device and each point radius is scaled relative to RSSI value.}
% \end{figure}

% \begin{figure} [ht]
% %    \begin{subfigure}
%         \includegraphics[width=5.35cm]{parts/multiple_device_measurements_anon.png}
%         \label{image:multiple_devices}
% %    \end{subfigure}
%     \hfill
% %    \begin{subfigure}
%         \includegraphics[width=2.65cm]{parts/commute_measurements_anon.png}
%         \label{image:commute_measurements}
% %    \end{subfigure}
%     \caption{Server-generated overlays of remote RSSI measurements with a known transmitter using multiple {\Sitara}s (left) and remote RSSI measurements crowd-sourced during a commute (right) (yellow blocks are used to conceal identifying information during review)}
%     \label{image:crowd-source}
% \end{figure}

We present these examples to demonstrate the utility of the {\Sitara} for conducting distributed spectrum measurement experiments. In Section \ref{section:discussion}, we  discuss ways we can adapt our platform to accommodate much more sophisticated experiments.

\subsection{Server-side Processing with GNU Radio}

In this application, we report the capability of the {\Sitara} to capture IQ samples from an over-the-air FSK transmission, upload the raw data to a remote server and recover the original message by demodulating the signal in software using GNU Radio. While this may superficially appear to be an unnecessarily complex method of performing a relatively simple demodulation, we present this as a proof of concept and later discuss the compelling implications when extended to future areas of research.

\begin{figure}[ht]
\includegraphics[width=8.4cm]{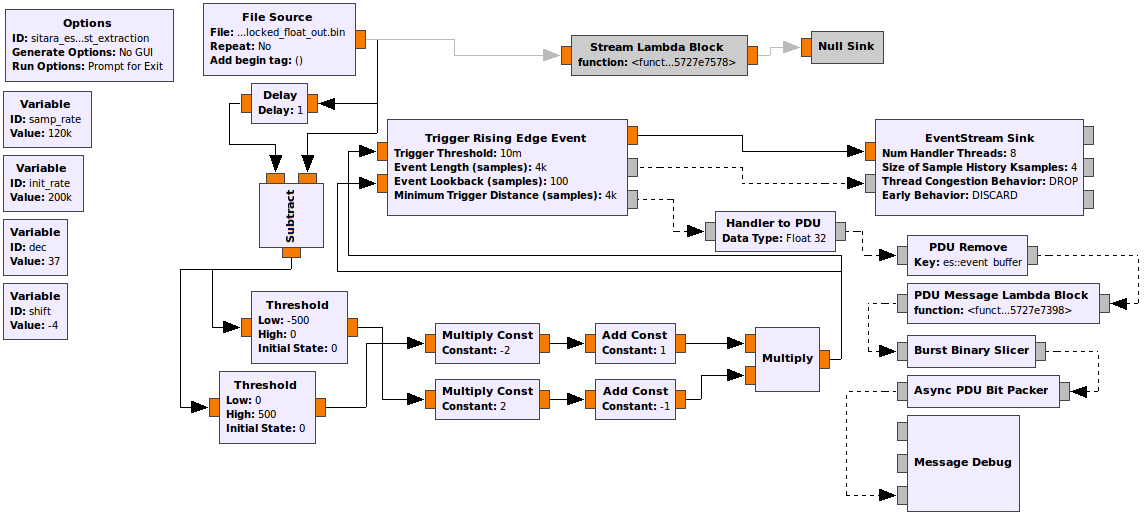}
\caption{A GNU Radio flow-graph used to generate the 2-FSK demodulator script}
\label{image:gnuradio}
\end{figure}

\begin{figure}[ht]
\includegraphics[width=8.4cm]{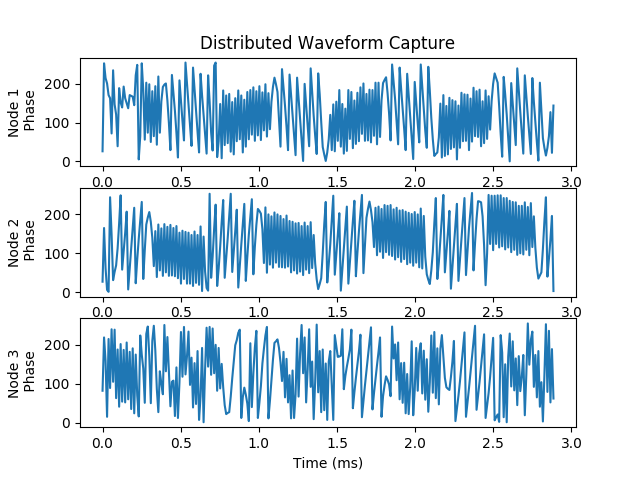}
\caption{Waveform phase measurements obtained from a triggered waveform capture using three {\Sitara}s}
\label{image:multi_signals}
\end{figure}

The experimental setup uses two {\Sitara}s, one acting as the transmitter, the other as the receiver. The user gives the transmitter a transmit (see Table \ref{table:commands}) command which directs the {\Sitara}'s radio to send a short message consisting of a preamble, sync word and user-defined payload, using the CC1200's native 2-FSK modulation format. The receiving {\Sitara}, instead of demodulating the signal directly in hardware, is configured to perform a triggered capture command while a carrier signal is detected. Once the sample capture is complete, the IQ samples, consisting of 16 bit integers, are uploaded to the server where further processing occurs. In this case, the measurements are converted to floating point for ease of use with our GNU Radio flow-graph. A plot of the angle samples can be seen in Fig.~\ref{image:multi_signals}. A Python script is generated from the flow-graph which extracts the payload message from the signal after synchronizing on the preamble and sync-word. A screen capture of the GNU Radio flow graph is shown in Fig.~\ref{image:gnuradio}. 
Beginning with an initial estimate for sampling rate and frequency offset, an iterative approach is used for symbol timing recover. As this experiment was only for demonstration purposes we opted for a naive implementation for the software receiver.

% \begin{figure}[ht]
% \includegraphics[width=8.4cm]{parts/gnu_radio_signal.png}
% \caption{A GNU Radio plot of the angle measurements obtained from a waveform captured by a {\Sitara} and uploaded to the server; Evident are the negative and positive frequency deviations of the 2-FSK signal}
% \label{image:signal}
% \end{figure}

Although only one captured signal was needed to demodulate the incoming signal in this case, using multiple {\Sitara}s, several copies of the same waveform captured at different geographical locations can be uploaded to the server for more sophisticated signal processing. Fig.~\ref{image:multi_signals} presents example snapshots from such a capture event. Here the waveforms do not appear to be synchronized, but the relative timing information can be recovered using the capture timestamp and correlating between waveforms. This example demonstrates the {\Sitara}'s capability to capture an RF signal and upload the data to the server for cloud-based reception.  We want to enable experimentation in coordinated mobile multi-antenna reception, using centralized cloud computing, to be able to separate signal from interference and demodulate signals that may not be recoverable from any one receiver alone.  Such experiments have been demonstrated to be powerful with static receivers \cite{charm}, and {\Sitara} can enable such experiments with crowd-sourced mobile endpoints.

%% file: parts/related_work.tex
\section{Related work}
We briefly compared {\Sitara} with other SDRs in Section~\ref{intro}. In this section, we expand the comparison to examine their capabilities and suitability for our application. Table \ref{table:platforms} provides a side-by-side comparison of size, control interface, power source and approximate price for a number of SDRs. 

\begin{table}[h]
\captionsetup{position=below, skip=10pt}
\begin{tabular}{p{1.5cm} | c |  c |  c | c}
SDR & Size (mm) & Interface & Power & Price\\
%\cline{1-5}
\hline \hline
\small USRP & 133x68x32 & USB/Eth & batt. & \$3200\\
\small PSDR & 110x70x30 & USB & batt. & \$500\\
\small HackRF & 124x80x18 & USB & USB & \$300\\
\small LimeSDR & 100x60 & USB & USB & \$300\\
\small PlutoSDR & 117x79x24 & USB & USB & \$100\\
\small RTL-SDR & 69x27x13 & USB & USB & \$20\\
\small Charm & 89x57x25 & Eth & PoE & n/a\\
\small {\Sitara} & 70x50x20 & BT 5 & batt. & \$38\\
\hline
\end{tabular}
\caption{Price and portability comparison of alternative SDR platforms}
\label{table:platforms}
\end{table}

We immediately see clear distinctions between these devices. {\Sitara} is the only device that allows data transfer and control without external connections. High speed Bluetooth 5 was not available during development of most of the devices in the list, and prior Bluetooth versions offered either high throughput or energy efficiency, but not both~\cite{blue}. Other wireless solutions such as WiFi-direct provided high throughput but consume too much power~\cite{carroll} and can be inconvenient to configure. Beyond power consumption, a direct WiFi connection would not be suitable for our application because it would be disruptive to the user's smartphone WiFi connection and require the cellular connection for access to the server. {\Sitara} is also among the smallest of the devices listed, second only to the RTL-SDR, which cannot function without a USB host. The PSDR is noteworthy in that it is relatively small, battery powered and can perform many functions as a standalone device; however, its hardware is limited to shortwave radio operation, up to about 30 MHz~\cite{ColtonSDR}. It does contain a GPS sensor, but it provides no mechanism to conveniently aggregate and monitor signals over a wide area for spectrum monitoring functions. The other devices could be used to perform spectrum monitoring functions and signal processing but are not suitable for wide scale, mobile experiments.

Beyond a comparison of sensor devices and software radios, we can also compare our system to other recent distributed software defined radio and spectrum sensing efforts. Two examples of distributed sensing systems which offer similarities are the Electrosense~\cite{electrosense} and Charm~\cite{charm} systems. Each of these employ stationary sensors and gateways which upload IQ measurements to a server. Electrosense is specifically designed for crowd-sourced spectrum monitoring, offering an open API and web-based frontend with a convenient waterfall display to visualize measurement data. Charm similarly samples RF spectrum to upload to a server, but instead of being used for spectrum monitoring specifically, it uses similar concepts to coherently combine distributed sensor measurements to recover degraded RF signals~\cite{charm}. 
% This would be a good place to add a sentence about what Pocket SDR enables in addition to these two.

%% file: parts/conclusion.tex
\section{Discussion}
\label{section:discussion}

\subsection{What is next?}
 We plan to expand our framework to test more sophisticated localization algorithms and develop propagation models enabled by our rapid data gathering capability across many sensors. Hundreds of participants can accumulate millions of pair-wise propagation measurements between different transmitting and receiving nodes to generate new highly accurate trained models. Extending the server-side GNU Radio application, we can apply {\Sitara} to experiments in distributed, coordinated signal processing as proposed by others~\cite{electrosense, charm, chorus}. Our system can be adapted to operate as a mesh network with other devices, perhaps as a limited fallback for cellular networks. By adapting the network topology and allowing a single mobile gateway to act as a central device to multiple {\Sitara}s, and improving the ability to synchronize the clock oscillator, we may leverage diversity gain and perform experiments in distributed multiple input multiple output (MIMO) and coherent combining at each node.

\subsection{When should {\Sitara} not be used?}
Having presented a number of applicable use cases for our system, there remain areas where the {\Sitara} is not particularly well suited. The CC1200 transceiver and RF front end limit the frequencies in which {\Sitara} can operate. Additionally, the maximum sampling rate and Bluetooth throughput constrain the SDR to narrowband operation. This is a direct consequence of the design choices made to minimize cost and maximize portability via a wireless back-channel. A more sophisticated receiver would require an FPGA or another ASIC with a much higher clock frequency and a delicate analog RF front-end. This would not only substantially reduce battery life, but also increase device cost. Our platform is designed to operate in ISM bands where transmission is permissible and operating frequency is inherently restricted; it is not intended to be a wideband receiver. Despite these limitations, there still exist many applications where {\Sitara} would be a suitable test platform~\cite{Hattab2018}. \\%A final limitation of the {\Sitara} is synchronization. The clock stability and interrupt timing resolution would present challenges for use cases requiring high clock accuracy such as time-of-arrival (TOA) ranging or distributed large-scale MIMO system.

\subsection{Challenges of Crowd-Sourcing}
A final topic not previously addressed in this paper but vital to the success of any crowd-sourced operation is that of user participation. A valid concern relating to the deployment of any crowd-sourced system is privacy. Increased scrutiny about how user data is collected, and potential for misuse, would necessarily impact how this system could be deployed. Does the information collected guarantee an appropriate level of anonymity? What techniques are possible which could conceal the precise location of participants but still provide enough information to capture useful information for sensing applications? Another question of equal if not greater importance is that of recruitment. What incentive do individuals have to participate in such a system--especially if participation could present a burden in any way to daily activities? If we wish to use the system to locate unauthorized transmitters then we could seek to determine the actual economic cost of RF spectrum abuse in order to assign a value to this data. Once the cost is established, we would allocate a commensurate reward to individuals that help locate an offender. These questions of privacy and incentive are beyond the scope of the work presented here, but we intend to explore these ideas in future work.

\section{Conclusion}
\label{section:conclusion}
The absence of viable options for large scale, coordinated, crowd-sourced spectrum sensing catalyzed our development of {\Sitara}, which we present here. We characterize the system and highlight its advantages for distributed spectrum measurement activities. We promote our design based on its merits as follows:
\begin{itemize}
    \item Energy efficiency, with a battery life lasting up to one week --- sufficient for a broad range of experiments.
    \item An inexpensive, compact form-factor including a wireless back-haul, offering an ideal solution for mobile, crowd-sourced scenarios. 
    \item Capability of local, manual or automated, and remote operation of sensors within a network distributed across a wide geographical area.
    \item SDR capabilities to measure complex temporal and spatial RF interactions. 
\end{itemize}
We showcase the {\Sitara}'s capabilities in real-world scenarios and evaluate its performance. The {\Sitara} 
%exceeds the initial requirements laid out for our test platform and becomes 
is a valuable open-source resource for research in distributed software-defined radio sensing.

%% file: main.bbl
\begin{thebibliography}{10}

\bibitem{arm}
{\sc ARM}.
\newblock {GNU ARM} embedded toolchain.
\newblock [Online], June 2018.

\bibitem{blue}
{\sc Bluetooth SIG}.
\newblock {\em Bluetooth Core Specification}, December 2016.
\newblock v5.

\bibitem{carroll}
{\sc Carroll, A., and Heiser, G.}
\newblock An analysis of power consumption in a smartphone.
\newblock In {\em Proceedings of the 2010 USENIX Conference on USENIX Annual
  Technical Conference\/} (Berkeley, CA, USA, 2010), USENIXATC'10, USENIX
  Association, pp.~21--21.

\bibitem{wireless_localization}
{\sc Chan, E. C.~L.}
\newblock {\em Introduction to wireless localization : with iPhone SDK
  examples}.
\newblock 2012.

\bibitem{gunicorn}
{\sc Chesneau, B.}
\newblock {Gunicorn WSGI Server}.
\newblock [Online], October 2017.
\newblock v19.9.0.

\bibitem{ColtonSDR}
{\sc Colton, M.}
\newblock Portablesdr.
\newblock [Online], June 2014.

\bibitem{electrosense}
{\sc den Bergh, B.~V., Giustiniano, D., Cordob\'es, H., Fuchs, M., Pollin, R.
  C.-P.~S., Rajendran, S., and Lenders, V.}
\newblock Electrosense: Crowsourcing spectrum monitoring.
\newblock {\em IEEE International Symposium on Dynamic Spectrum Access Networks
  (DySPAN 2017)\/} (March 2017), 1--2.

\bibitem{charm}
{\sc Dongare, A., Narayanan, R., Gadre, A., Luong, A., Balanuta, A., Kumar, S.,
  Iannucci, B., and Rowe, A.}
\newblock Charm: Exploiting geographical diversity through coherent combining
  in low-power wide-area networks.
\newblock In {\em Proceedings of the 17th ACM/IEEE International Conference on
  Information Processing in Sensor Networks\/} (Piscataway, NJ, USA, 2018),
  IPSN '18, IEEE Press, pp.~60--71.

\bibitem{ettus}
{\sc Ettus, M.}
\newblock {U}niversal {S}oftware {R}adio {P}eripheral.
\newblock [Online].

\bibitem{e312}
{\sc Ettus Research}.
\newblock {\em Getting Started with the Ettus Research USRP E312 SDR}, March
  2016.
\newblock Rev 1.

\bibitem{sigmf}
{\sc {GNU Radio Foundation}}.
\newblock {\em Signal Metadata Format Specification}, 7 2018.

\bibitem{chorus}
{\sc Hamed, E., Rahul, H., and Partov, B.}
\newblock Chorus: Truly distributed distributed-mimo.
\newblock In {\em Proceedings of the 2018 Conference of the ACM Special
  Interest Group on Data Communication\/} (New York, NY, USA, 2018), SIGCOMM
  '18, ACM, pp.~461--475.

\bibitem{Hattab2018}
{\sc {Hattab}, G., and {Cabric}, D.}
\newblock Spectrum sharing protocols based on ultra-narrowband communications
  for unlicensed massive iot.
\newblock In {\em 2018 IEEE International Symposium on Dynamic Spectrum Access
  Networks (DySPAN)\/} (Oct 2018), pp.~1--10.

\bibitem{synch}
{\sc He, J., Cheng, P., Shi, L., and Chen, J.}
\newblock Time synchronization for random mobile sensor networks.
\newblock vol.~63, pp.~2712--2717.

\bibitem{luong2016rss}
{\sc Luong, A., Abrar, A.~S., Schmid, T., and Patwari, N.}
\newblock Rss step size: 1 db is not enough!
\newblock In {\em Proceedings of the 3rd Workshop on Hot Topics in Wireless\/}
  (2016), ACM, pp.~17--21.

\bibitem{Luong:2018:STF:3207947.3207959}
{\sc Luong, A., Hillyard, P., Abrar, A.~S., Che, C., Rowe, A., Schmid, T., and
  Patwari, N.}
\newblock A stitch in time and frequency synchronization saves bandwidth.
\newblock In {\em Proceedings of the 17th ACM/IEEE International Conference on
  Information Processing in Sensor Networks\/} (Piscataway, NJ, USA, 2018),
  IPSN '18, IEEE Press, pp.~96--107.

\bibitem{propagation}
{\sc {Meng}, Y.~S., {Lee}, Y.~H., and {Ng}, B.~C.}
\newblock Empirical near ground path loss modeling in a forest at vhf and uhf
  bands.
\newblock {\em IEEE Transactions on Antennas and Propagation 57}, 5 (May 2009),
  1461--1468.

\bibitem{fpga}
{\sc {MicroSemi}}.
\newblock {\em Energy Efficient Digital Frontend Designs With PolarFire FPGAs
  For Small Cells}, 2 2008.

\bibitem{bt_antenna}
{\sc MOLEX}.
\newblock {\em 2.4 GHz SMD On-Ground MID Chip Antenna 0479480001}, June 2018.
\newblock rev 4.

\bibitem{nrf}
{\sc Nordic Semiconductor}.
\newblock {\em nRF52840 Product Specification}, March 2018.
\newblock v1.0.

\bibitem{scos}
{\sc NTIA}.
\newblock {\em IEEE 802.22.3 Spectrum Characterization and Occupancy Sensing},
  1 2018.
\newblock v0.0.2.

\bibitem{flask}
{\sc Ronacher, A.}
\newblock Flask.
\newblock [Online], May 2018.
\newblock v1.0.2.

\bibitem{rtlSDR}
{\sc {RTL-SDR}}.
\newblock {RTL-SDR}.
\newblock [Online].

\bibitem{shafi2017}
{\sc Shafi, M., Molisch, A.~F., Smith, P.~J., Haustein, T., Zhu, P., Silva,
  P.~D., Tufvesson, F., Benjebbour, A., and Wunder, G.}
\newblock 5g: A tutorial overview of standards, trials, challenges, deployment,
  and practice.
\newblock {\em IEEE Journal on Selected Areas in Communications 35}, 6 (June
  2017), 1201--1221.

\bibitem{sharma2017dynamic}
{\sc Sharma, S.~K., Bogale, T.~E., Le, L.~B., Chatzinotas, S., Wang, X., and
  Ottersten, B.}
\newblock Dynamic spectrum sharing in {5G} wireless networks with full-duplex
  technology: Recent advances and research challenges.
\newblock {\em IEEE Communications Surveys Tutorials 20}, 1 (Firstquarter
  2018), 674--707.

\bibitem{limeSDR}
{\sc Tamosevicius, Z., and Kiela, K.}
\newblock {LimeSDR}.
\newblock [Online].

\bibitem{cc1200}
{\sc Texas Instruments}.
\newblock {\em CC1200 Low-Power, High-Performance RF Transceiver}, July 2013.

\bibitem{wang2014cellular}
{\sc Wang, C., Haider, F., Gao, X., You, X., Yang, Y., Yuan, D., Aggoune,
  H.~M., Haas, H., Fletcher, S., and Hepsaydir, E.}
\newblock Cellular architecture and key technologies for {5G} wireless
  communication networks.
\newblock {\em IEEE Communications Magazine 52}, 2 (February 2014), 122--130.

\bibitem{improved_wifi}
{\sc Xue, W., qiu, w., Hua, X., and Yu, K.}
\newblock Improved wi-fi rssi measurement for indoor localization.
\newblock {\em IEEE Sensors Journal PP\/} (01 2017), 1--1.

\bibitem{antenna}
{\sc YAGEO}.
\newblock {\em PCB type antenna ANTX100P001BWPEN3}, June 2014.
\newblock v0.0.

\bibitem{fingerprint}
{\sc Zhou, M., Tian, Z., Xu, K., Yu, X., and Wu, H.}
\newblock Theoretical entropy assessment of fingerprint-based wi-fi
  localization accuracy.
\newblock {\em Expert Systems with Applications 40}, 15 (2013), 6136 -- 6149.

\end{thebibliography}
